\documentclass[prb,twocolumn,showpacs,preprintnumbers,amsmath,amssymb]{revtex4}
\usepackage{graphicx}
\usepackage{epsfig}

\newcommand{\be}{\begin{eqnarray}}
\newcommand{\ee}{\end{eqnarray}}

\newcommand{\bi}{\begin{itemize}}
\newcommand{\ei}{\end{itemize}}

\begin{document}
\title{Interaction between ionic 
lattices and superconducting condensates}
\author{Pavel Lipavsk\'y$^{1,2}$, Klaus Morawetz$^{3,4}$, 
Jan Kol\'a\v cek$^2$, Ernst Helmut Brandt$^5$ and Michael Schreiber$^3$}
\affiliation{
$^1$ Faculty of Mathematics and Physics, Charles University, 
Ke Karlovu 3, 12116 Prague 2, Czech Republic}
\affiliation{
$^2$Institute of Physics, Academy of Sciences, 
Cukrovarnick\'a 10, 16253 Prague 6, Czech Republic}
\affiliation{
$^3$Institute of Physics, Chemnitz University of Technology, 
09107 Chemnitz, Germany}
\affiliation{
$^4$Max-Planck-Institute for the Physics of Complex
Systems, Noethnitzer Str. 38, 01187 Dresden, Germany}
\affiliation{
$^5$Max-Planck-Institute for Metals Research,
         D-70506 Stuttgart, Germany}
\begin{abstract}
The interaction of the ionic lattice with the superconducting
condensate is treated in terms of the electrostatic force in
superconductors. It is shown that this force is similar but not
identical to the force suggested by the volume difference of
the normal and superconducting states. The BCS theory shows 
larger deviations than the two-fluid model.
\end{abstract}

\maketitle
\section{Introduction}
The theory of deformable superconductors deals either with 
effects of the lattice deformations on the superconducting 
condensate or with deformations of the crystal lattice 
driven by the inhomogeneous superconducting condensate. For
example, a lattice deformation around a dislocation pins 
a vortex.\cite{KB67,L68} In contrary, forces generated by
supercurrents contribute to the  
magnetostriction\cite{IHNKK93,NKGBGSM97,CIL05} and 
a condensate depletion at the vortex core deforms the lattice 
so strongly that a significant renormalization of the vortex 
mass has been predicted.\cite{Sim91,DSim92,C94,CK03} In some 
cases one cannot say which of the two effects are dominant. 
This happens, for instance, if the 
ionic lattice deformation influences the structure or 
orientation of the Abrikosov vortex lattice.\cite{KBMD95}

The free energy describing deformable superconductors has to
include at least three parts. The first part is the elastic 
energy of deformations. Its structure and parametrization have well been 
established for a long time.\cite{LL75} The second part
is the magnetic energy and the energy of superconducting 
condensation. This part can be covered on different levels. 
Here we will refer to the Ginzburg-Landau (GL) theory\cite{GL50} 
employed in the majority of the above mentioned studies. 
The third part is a cross term which describes the mutual 
effect of deformations and the condensate. In this paper 
we focus on this interaction term.

In the phenomenological approach put forward by Kramer and 
Bauer\cite{KB67} the interaction of the lattice and the condensate 
is described by a local product of the lattice density with 
the density of superconducting electrons. If the lattice is
modeled by an isotropic deformable medium and the interaction
is assumed to be local, the interaction energy of Kramer
and Bauer is the only one compatible with the system 
symmetry. Indeed, the condensate density is a scalar which can
interact only with another scalar. The only isotropic scalar 
quantity linear in the deformation is the trace of the strain 
tensor, which is proportional to the ionic density. Of course,
one can construct more elaborate scalars within non-linear
terms but these are higher-order corrections.

Two modifications are at hand. First, one can take into
account that the real lattice is never isotropic. Even in
simplest lattices of elementary metals the shear rigidity
depends on the orientation of the deformation with respect 
to the crystal axes. In the anisotropic crystal, there are 
two scalar quantities linear in the shear deformation. 
A corresponding anisotropic generalization of the interaction 
between the condensate and the ionic lattice has been discussed
by Kogan {\em et al}.\cite{KBMD95}

Second, one can go beyond the local approximation. The reason
for such a step is the following. The interaction in the local 
approximation is justified only for neutral systems.\cite{LL75} 
The superconducting condensate, however, drives the system
out of neutrality inducing the electrostatic potential known 
as the Bernoulli potential.\cite{BK68,MB71} The non-local
interaction mediated by the Bernoulli potential in the
bulk of the superconductor has been discussed in our 
previous paper.\cite{LMKB07} 

Another contribution to the charge transfer induced by the
condensate is the surface dipole.\cite{LMKMBSa04} While all 
the above mentioned interactions result in a force density
acting in the bulk of the crystal, the surface dipole yields 
the force which acts as an external pressure imposed on the
surface. As far as we know, the surface dipole has never been 
discussed within the theory of deformable superconductors.
In this paper we want to fill this gap.

The paper is organized as follows. In section II we show 
that the surface dipole determines changes of the crystal
volume during its transition from the normal to the 
superconducting state. To this end we first introduce the 
basic concept in Sec.~II.A and derive the coefficient of 
the local interaction from the pressure dependence of the 
condensation energy at zero temperature in Sec.~II.B. The 
result is compared with the force due to the surface dipole 
in Sec.~II.C. In section III we introduce the interaction 
mediated by the Bernoulli potential. We first derive a
formula for the coefficient of the local interaction. In 
Secs.~III.A and III.B we evaluate the interaction coefficient 
for moderately strong and weak coupling superconductors.
In section IV we discuss differences and conclude.

\section{Local approach}

In their pioneering study Kramer and Bauer\cite{KB67} 
proposed to deduce the interaction strength from the
pressure dependence of the critical magnetic field 
$B_{\rm c}$. Since experimental data for this parametrization 
are conveniently found in literature, this approach has been 
employed by other authors too. 

In this section we provide a derivation of the local
interaction of Kramer and Bauer within the GL
picture of the superconductor. The presented approach is based 
on papers by {\v S}im{\'a}nek\cite{Sim91} and Hake.\cite{H68}

\subsection{Phenomenological force}

The volume of a metal changes at the phase transition from 
$V_{\rm n}$ in the normal state to $V_{\rm s}$ in the superconducting 
state.\cite{S52,H68} This change is described by a relative 
change $\alpha$ of the specific volume defined as
\begin{equation}
V_{\rm n}-V_{\rm s}=\alpha\,V_{\rm s}.
\label{e1}
\end{equation}

If the specific volume becomes inhomogeneous, the crystal 
has regions requiring different distances of neighboring atoms.
This leads to internal stresses which
can be expressed via an effective force density \cite{LL75}
\begin{equation}
{\bf F}_{\rm ph}=K\,\nabla \alpha,
\label{e2p1}
\end{equation}
where $K$ is the modulus of hydrostatic compression or simply
the bulk modulus. It is defined as the inverse of the relative 
volume change with respect to the pressure
\begin{equation}
{1\over K}=-{1\over V}
{\partial V\over\partial p}.
\label{e7}
\end{equation}

The temperature dependence of $\alpha$ is similar to the
temperature dependence of the superconducting fraction
\begin{equation}
{\alpha\over\alpha_0}\approx {|\psi|^2\over|\psi_0|^2},
\label{e2p2}
\end{equation}
where subscript zero denotes the zero temperature value.
{\v S}im{\'a}nek\cite{Sim91,DSim92} and Coffey\cite{C94,CK03} 
use the BCS gap $\Delta$,
\begin{equation}
{\alpha\over\alpha_0}
\approx{|\Delta|^2\over|\Delta_0|^2},
\label{e2p2a}
\end{equation}
while other authors prefer the GL 
function $\psi$. We will restrict our attention to the 
vicinity of the critical temperature, where both forms
are equivalent since the BCS gap and the GL function
are linearly proportional to each other.\cite{Gor59}

Relation (\ref{e2p2}) is the central approximation in the 
phenomenological theory of deformable superconductors. 
Assuming that in the normal state the system is homogeneous,
$\alpha_0$ and $\psi_0$ are constants. With the GL function
normalized to the density of pair-able electrons
$2|\psi_0|^2=n$, we obtain
\begin{equation}
{\bf F}_{\rm ph}={2\over n}\,\alpha_0\,K\,\nabla |\psi|^2
\label{e2}
\end{equation}
which we will use in our discussion.

\subsection{Difference of normal and superconducting volume}

Now we link the relative change of the specific volume $\alpha$ 
to the pressure dependence of the condensation energy 
$\varepsilon_{\rm con}$. We follow the derivation of Hake.\cite{H68} 

The volume of the sample is the pressure derivative at fixed 
temperature of the Gibbs free energy 
\begin{equation}
V=\left({\partial G\over\partial p}\right)_T.
\label{e3}
\end{equation}
At zero magnetic field and zero temperature, the 
free energy of the normal state $G_{\rm n}$ is higher than 
the superconducting free energy $G_{\rm s}$ by the condensation energy
\begin{equation}
G_{\rm n}-G_{\rm s}=V_{\rm s}\varepsilon_{\rm con}.
\label{e4}
\end{equation}

Due to the complete expulsion of the magnetic field from 
type-I superconductors, the condensation energy is conveniently 
observed via the critical magnetic field at zero temperature 
\begin{equation}
\varepsilon_{\rm con}={B_0^2\over 2\mu_0}.
\label{e12}
\end{equation}
In his study, Hake expresses all thermodynamical relations
exclusively in terms of the critical magnetic field. Here
we prefer to use the condensation energy.

>From the pressure derivative of the relation (\ref{e4}) it follows
\begin{equation}
V_{\rm n}-V_{\rm s}=V_{\rm s}
{\partial \varepsilon_{\rm con}\over\partial p}+
\varepsilon_{\rm con}
{\partial V_{\rm s}\over\partial p}.
\label{e5}
\end{equation}
Comparing the thermodynamical relation (\ref{e5}) with 
the definition (\ref{e1}) we obtain the coefficient $\alpha_0$ 
in terms of the condensation energy
\begin{equation}
\alpha_0={\partial \varepsilon_{\rm con}\over\partial p}+
\varepsilon_{\rm con}
{1\over V_{\rm s}}
{\partial V_{\rm s}\over\partial p}.
\label{e6}
\end{equation}
In terms of the bulk modulus (\ref{e7}) we have
\begin{equation}
\alpha_0={\partial \varepsilon_{\rm con}\over\partial p}-
{\varepsilon_{\rm con}\over K}.
\label{e9}
\end{equation}

The force (\ref{e2}) depends on the product $\alpha_0 K$, 
it is thus advantageous to introduce the 
inverse bulk modulus also into the first term of 
Eq.~(\ref{e9}). For simplicity we consider hydrostatic 
pressure and conventional superconductors
with isotropic structure. In this case we can express 
the pressure dependence of the density of the condensation 
energy $\varepsilon_{\rm con}$ via its dependence on the 
electron density,
\begin{equation}
{\partial \varepsilon_{\rm con}\over\partial p}=
{\partial \varepsilon_{\rm con}\over\partial n}
{\partial n\over\partial p}.
\label{e10}
\end{equation}
Since the number of electrons $N$ does not change, we
can express the bulk compressibility via the
change of the density $n=N/V$ as
\begin{equation}
{1\over K}={1\over n}
{\partial n\over\partial p}.
\label{e8}
\end{equation}
Using relation (\ref{e10}) and the bulk modulus
(\ref{e8}) in equation (\ref{e9}) we get
\begin{equation}
\alpha_0 K=n
{\partial \varepsilon_{\rm con}\over\partial n}-
\varepsilon_{\rm con}.
\label{e13}
\end{equation}
The density derivative is taken under the condition of
charge neutrality, i.e., the lattice density changes 
with the electron density.

The phenomenological force (\ref{e2}) according to 
relation (\ref{e13}) thus reads
\begin{equation}
{\bf F}_{\rm ph}=2 \left(
{\partial \varepsilon_{\rm con}\over\partial n}-
{\varepsilon_{\rm con}\over n}\right)\nabla|\psi|^2.
\label{e14}
\end{equation}
We will compare this form with the electrostatic force resulting from 
a surface dipole.

\subsection{Compression via the surface dipole}
At the surface of a metal the electrostatic potential
rises by few Volts from its vacuum value to the value 
deep in the metal.\cite{KW96} This increase is spread partly 
outside the region occupied by ions, typically on the 
scale of the tunneling length of electrons in the potential
barrier given by the work function. A part of the barrier is 
located inside the metal on the scale of
the Thomas-Fermi screening length. Both scales are
of the order of \AA ngstr\o ms making the potential
step very sharp. This sharp step is called the surface 
dipole.

The surface dipole naturally depends on the temperature.
Moreover, when the metal undergoes a transition to the 
superconducting state, the temperature dependence of the
surface dipole changes. Briefly, the superconducting
condensate affects the surface dipole.\cite{LMKMBSa04}
Let us denote this additional potential near the surface
as $\varphi_T$.

An amplitude $\varphi_T(0)-\varphi_T(\infty)$ 
of the additional potential step follows from the 
Budd-Vannimenus theorem as\cite{LMKMBSa04}
\begin{equation}
\rho_{\rm lat}\left [\varphi_T(0)-\varphi_T(\infty)\right]=f-
n{\partial f\over\partial n},
\label{e15a}
\end{equation}
where $f=f_{\rm s}-f_{\rm n}$ is the free-energy density 
by which the superconducting state differs from the normal 
state and where $\rho_{\rm lat}$ is the charge density of the 
ionic lattice.
We assume a superconductor which fills the 
half-space $x>0$.

The additional potential exerts an electrostatic 
force density on the ionic lattice
\begin{equation}
{\bf F}_T=-\rho_{\rm lat}\nabla\varphi_T.
\label{e15b}
\end{equation}
The integral of this force density across the surface
region corresponds to an effective pressure on the lattice
\begin{equation}
p_T=\int\limits_0^\infty dx F_T^x=
\rho_{\rm lat}\left[\varphi_T(0)-\varphi_T(\infty)\right],
\label{e15c}
\end{equation}
which changes the volume of the crystal by
\begin{equation}
\tilde\alpha \,V={\partial V\over \partial p}p_T.
\label{e15d}
\end{equation}
Clearly, the surface dipole contributes to the relative 
change of the specific volume $\tilde\alpha$. From (\ref{e15d})
we find
\begin{equation}
\tilde\alpha={1\over K}\rho_{\rm lat}\left[\varphi_T(0)-
\varphi_T(\infty)\right].
\label{e15e}
\end{equation}

At zero temperature $f=-\varepsilon_{\rm con}$,
therefore from (\ref{e15a}) it follows
\begin{equation}
\rho_{\rm lat}\left[\varphi_0(0)-\varphi_0(\infty)\right]=
n{\partial\varepsilon_{\rm con}\over\partial n}-
\varepsilon_{\rm con}.
\label{e15f}
\end{equation}
The dipole-induced volume change at zero temperature
thus reads
\begin{equation}
\tilde\alpha_0={1\over K}\left(n{\partial\varepsilon_{\rm con}
\over\partial n}-\varepsilon_{\rm con}\right).
\label{e15g}
\end{equation}
Comparing (\ref{e15g}) with (\ref{e13}) one can see that
the volume change is fully induced by the surface dipole
\begin{equation}
\alpha_0=\tilde\alpha_0.
\label{e15h}
\end{equation}

The fact that the volume change is driven by the surface 
dipole shows that one should be cautious using the relative 
change of the specific volume $\alpha_0$ as a coefficient 
of the interaction between the ionic lattice and the 
condensate. 

Studies of the electrostatic potential in superconductors
have shown that the bulk and surface potentials are of
different nature and are covered by distinct theories.
We note that these theories are experimentally verified.
The surface potential including the surface dipole has 
been observed by Morris and Brown via the Kelvin capacitive
pickup.\cite{MB71,LMKMBSa04} The internal charge transfer 
caused by the bulk electrostatic potential in the vortex
core has been observed by Kumagai, Nozaki and Matsuda 
via the nuclear magnetic resonance.\cite{KNM01,LKMB02}

\section{Electrostatic force on ions}
According to the Hellmann-Feynman theorem, electrons 
act on ions exclusively via the electrostatic 
force.\footnote{One should be cautious about using this 
general argument. The electronic density is modulated
by the ionic potential which results in large local
electric fields. Their total force on ions is nontrivial, 
in particular, when the lattice is strained. Formula 
(\ref{e15}) omits all local contributions. We do not 
include these contributions in this paper and 
leave them for the future work.}
In this spirit, we expect the force density to be
of electrostatic nature,
\begin{equation}
{\bf F}_{\rm el}=-\rho_{\rm lat}\nabla\varphi,
\label{e15}
\end{equation}
where $\varphi$ is the electrostatic 
potential created by the superconducting electrons which
is conveniently
derived following the approach of Rickayzen.\cite{Ri69}
Since the system is in equilibrium, the Gibbs 
electrochemical potential $\mu$ for electrons is constant 
all over the sample. It is locally defined from the
density of free energy $f$ as
\begin{equation}
\mu=e\varphi+{\partial f\over\partial n}.
\label{e16}
\end{equation}
Following the customary choice in the theory of
superconductivity we set the electrochemical potential
to zero, $\mu=0$, therefore
\begin{equation}
\varphi=-{1\over e}{\partial f\over\partial n}.
\label{e17}
\end{equation}

The theory of the electrostatic potential has been derived 
under the assumption that the ion lattice is stiff and its
deformation is not included. A combination of both
effects has not been studied so far, therefore it is 
not yet clear how the density derivative in (\ref{e17}) 
is modified by lattice deformations. For simplicity we
assume that the density derivatives in (\ref{e17}) and in
(\ref{e15a}) are the same. This is the case if the 
pairing interaction has a purely electronic nature so that
the lattice density has no effect on the condensation 
energy. 

Formula (\ref{e17}) is quite general. It has been employed
by Rickayzen to evaluate the Bernoulli potential in
superconductors using the London theory supplemented by the
phenomenological temperature dependence of the superconducting
density. In the same paper\cite{Ri69} Rickayzen has used
formula (\ref{e17}) with the BCS free energy and recovered
the result of Adkins and Waldram.\cite{AW68} 

We use the free-energy density in the GL approximation 
\begin{eqnarray}
f&=&a\left(T-T_{\rm c}\right)|\psi|^2+{1\over 2}b|\psi|^4
\nonumber\\
&+&{1\over 2m^*}
\left|\left(-i\hbar\nabla-e^*{\bf A}\right)\psi\right|^2+
{1\over 2\mu_0}\left|{\bf B}_{\rm a}-\nabla\times{\bf A}
\right|^2,
\nonumber\\
\label{e18}
\end{eqnarray}
where $\bf A$ is the vector potential and ${\bf B}_{\rm a}$ 
is the applied magnetic field. The reader not familiar with
the GL theory is referred to the textbook\cite{Tinkham} of Tinkham.

The magnetic free energy (the last term of (\ref{e18})) 
does not depend on the electron 
density. For simplicity we also assume that the Cooper 
pair mass $m^*$ is independent of this density. Since the GL 
wave function $\psi$ and the vector potential $\bf A$ are 
independent variational fields, the density derivative of 
the free energy yields the electrostatic potential
\begin{equation}
\varphi=
{a\over e}{\partial T_{\rm c}\over\partial n}|\psi|^2-
{T-T_{\rm c}\over e}{\partial a\over\partial n}|\psi|^2-
{1\over 2e}{\partial b\over\partial n}|\psi|^4.
\label{e19}
\end{equation}

The nonlocal corrections discussed in 
Ref.~\onlinecite{LMKB07} are hidden in the second and 
third terms of (\ref{e19}). They can be made explicit 
using relations for material parameters $a$ and $b$,
e.g. (\ref{e23}), and the GL equation, which couples
nonlocal and nonlinear contributions. 

Here we restrict our attention to a close vicinity of 
the critical temperature, where all nonlocal and 
nonlinear contributions can be neglected.
Indeed, for $T\to T_{\rm c}$ the GL wave function 
vanishes $|\psi|^2 \propto T_{\rm c}-T$. In
lowest order in $T-T_{\rm c}$ we can neglect the second 
and the quartic term so that relation (\ref{e19})
simplifies to
\begin{equation}
\varphi={a\over e}{\partial T_{\rm c}\over\partial n}
|\psi|^2.
\label{e20}
\end{equation}

Now we are ready to evaluate the force acting on the
ionic lattice. The electrostatic force density (\ref{e15}) with
the electrostatic potential (\ref{e20}) reads
\begin{equation}
{\bf F}_{\rm el}=-{\rho_{\rm lat}\over e}\nabla \left(a
{\partial T_{\rm c}\over\partial n}|\psi|^2\right).
\label{e21}
\end{equation}
In first order in $T_{\rm c}-T$, gradients 
of material parameters $a$ and $\partial T_{\rm c}/
\partial n$ do not contribute, i.e.,
\begin{equation}
{\bf F}_{\rm el}=a\, n\,{\partial T_{\rm c}\over\partial n}
\nabla |\psi|^2.
\label{e22}
\end{equation}
We have used $\rho_{\rm lat}=-en$ demanded by the local 
charge neutrality.

With nonlocal and nonlinear corrections neglected, 
the electrostatic force (\ref{e22}) like the 
phenomenological force (\ref{e14}) are proportional 
to the gradient of the superconducting density $|\psi|^2$.
Our next aim is to compare the electrostatic coefficient 
$a n{\partial T_{\rm c}\over\partial n}$ with its
phenomenological precursor $2({\partial \varepsilon_{\rm con}\over
\partial n}-{\varepsilon_{\rm con}\over n})$. We will 
show that the relative values of these coefficients depend 
on the strength of the pairing interaction.

\subsection{Superconductors with moderately strong coupling}
To be able to compare the electrostatic force density (\ref{e22})
with the phenomenological force density (\ref{e2}), we need the GL 
parameter $a$ as function of the electron density $n$. 
For metals like niobium or lead, it is possible to use 
the asymptotic form of the two-fluid free energy of 
Gorter and Casimir \cite{GC34} giving\cite{LKMB01}
\begin{equation}
a=a_{\rm GC}={\gamma T_{\rm c}\over n}.
\label{e23}
\end{equation}
Here $\gamma$ is the linear coefficient of the specific 
heat.

The critical temperature $T_{\rm c}$ and the critical 
magnetic field $B_0$ at zero temperature are linked 
via the condensation energy. The two-fluid model 
yields\cite{B55}
\begin{equation}
{1\over 4}\gamma T_{\rm c}^2={B_0^2\over 2\mu_0}.
\label{e24}
\end{equation}
Within this approximation, the 
electrostatic force density (\ref{e22}) reads
\begin{equation}
{\bf F}_{\rm el}^{\rm GC}=2
\left({\partial\varepsilon_{\rm con}\over\partial n}-
{1\over 4}T_{\rm c}^2
{\partial\gamma\over\partial n}\right)\nabla|\psi|^2.
\label{e25}
\end{equation}
One can see that this is similar but not identical
to the phenomenological force density (\ref{e14}). In particular,
the same dominant term 
$\propto {\partial\varepsilon_{\rm con}/\partial n}$
results from both approaches.

We note that {\v S}im{\'a}nek and many of other authors 
use the approximation 
$\left|{\partial\varepsilon_{\rm con}\over\partial n}\right|
\gg\left|{\varepsilon_{\rm con}\over n}\right|$, i.e.,
they consider only the derivative in their formulas. 
Within this accuracy both formulas are equivalent.

\subsection{Superconductors with weak coupling}
Metals like aluminum have weak electron-phonon coupling
and one can use BCS relations. This approximation
results in a slightly different electrostatic force.

>From the BCS theory Gor'kov has obtained parameters of the 
GL theory.\cite{Gor59} The linear GL
coefficient reads
\begin{equation}
a=a_{\rm BCS}={6\pi^2k_{\rm B}T_{\rm c}\over 7\zeta(3)E_{\rm F}},
\label{e25a}
\end{equation}
where $E_{\rm F}$ is the Fermi energy. 
The Riemann Zeta function has the value $\zeta(3)=1.202$.

The BCS and the Gorter and Casimir approximations of $a$
can be related within the free electron model. 
The electron density determines the Fermi vector 
$k_{\rm F}=(3\pi^2n)^{1/3}$ in terms of which
$E_{\rm F}=\hbar^2k_{\rm F}^2/2m$. We also use 
$\gamma=(2/3)\pi^2k_{\rm B}^2N_0$, where 
$N_0=(1/4\pi^2)(2m/\hbar^2)k_{\rm F}$
is the single-spin density of states. Combining these
relations one finds that both values differ by 
a numerical factor
\begin{equation}
a_{\rm BCS}={12\over 7\zeta(3)}~
a_{\rm GC}=1.43~a_{\rm GC}.
\label{e25aa}
\end{equation}

The BCS relation
connecting the critical temperature with the
condensation energy defined via the magnetic field yields 
another numerical factor 
\begin{equation}
0.947\,{1\over 4}\gamma T_{\rm c}^2={B_0^2\over 2\mu_0}.
\label{e25b}
\end{equation}
In the two-fluid model we find $T_{\rm c}^{\rm GC}=
B_0\sqrt{2\gamma/\mu_0}$. The correction 
\be
T_{\rm c}^{\rm BCS}=
{T_{\rm c}^{\rm GC}\over \sqrt{0.947}}=1.028~T_{\rm c}^{\rm GC}
\label{ad}
\ee 
is
by an order of magnitude less important than the factor 1.43 
from relation (\ref{e25aa}), however.

>From (\ref{e25b}) we obtain $T_{\rm c}$ in terms of
$B_0$, which we use in the force density (\ref{e22}). With the 
BCS relation (\ref{e25a}), 
the electrostatic force density (\ref{e22}) reads
\begin{eqnarray}
{\bf F}_{\rm el}^{\rm BCS}&=&{24\over 7\zeta(3)}
\left({1\over 0.947}
{\partial\varepsilon_{\rm con}\over\partial n}-
{1\over 4}T_{\rm c}^2
{\partial\gamma\over\partial n}\right)\nabla|\psi|^2
\nonumber\\
&=&3.012~\left(
{\partial\varepsilon_{\rm con}\over\partial n}-0.947
{1\over 4}T_{\rm c}^2
{\partial\gamma\over\partial n}\right)\nabla|\psi|^2.
\label{e25c}
\end{eqnarray}
One can see that the dominant contribution is 
increased by slightly more than 50\%\ as compared to the 
phenomenological force density (\ref{e14}) and the Gorter-Casimir 
approximation (\ref{e25}). Compared to the Gorter-Casimir approximation 
the critical temperature $T_{\rm c}^{\rm GC}$ is replaced additionally 
by the BCS 
value $T_{\rm c}^{\rm BCS}$. It will thus be interesting 
to test the validity of the phenomenological force on materials of
rather different coupling strength.

\section{Discussion}
We have shown that the change of the volume during the
superconducting transition can be expressed as a compression
caused by the surface dipole. In analo\-gy we have used the 
internal electrostatic potential for the lattice deformation 
in the bulk. The force resulting from the electrostatic 
potential in the superconductor is similar but not identical 
to the phenomenological force suggested from the volume 
change.

For moderately coupled materials well described by the 
Gorter-Casimir two-fluid model, the phenomenological and 
the electrostatic forces have identical dominant terms 
$\propto\partial\varepsilon_{\rm con}/\partial n$. They 
differ in the correction terms only. For weakly coupled 
superconductors covered by the BCS theory, the dominant 
term is increased by nearly 50\%. We have not discussed the 
strongly coupled superconductors which do not obey any of 
these limits. One can expect that the dominant term is 
reduced.

It is a question whether the above derived small 
differences in the internal forces can be accessed by 
some of recent experiments. Among the physical phenomena
mentioned in the introduction, the magnetostriction offers 
the most sensitive experimental technique. Indeed, one can 
resolve even deformations driven by such small changes in 
magnetization as those caused by the de Haas-van Alphen 
effect with relative changes of the susceptibility of the
order of one in ten millions.\cite{GCh63} A superconductor
in the Meissner state is an ideal diamagnet with large but 
fixed magnetization. Small deviations from ideality appear
due to the penetration of the magnetic field into the surface. At
the same time, the magnetostriction combines the internal
forces with the surface dipole. To separate these two contributions,
it will be necessary to analyze how the deformation depends 
on the sample geometry.

The two approaches compared in this paper represent two 
extreme models. In the electrostatic approach all forces 
are attributed to a mean electric field. In the 
phenomenological model the system is treated as locally 
neutral which implies that all forces are attributed to 
bonds between ions. We expect that a realistic description 
requires to combine both approaches. 

\medskip
This work was supported by research plans
MSM 0021620834 and No. AVOZ10100521, by grants
GA\v{C}R 202/07/0597 and GAAV 100100712, by
DAAD and by European ESF program AQDJJ.

\bibliography{bose,delay2,delay3,gdr,genn,chaos,kmsr,kmsr1,kmsr2,kmsr3,kmsr4,kmsr5,kmsr6,kmsr7,micha,refer,sem1,sem2,sem3,short,spin,spin1,solid,deform}

\end{document}